\documentclass[twocolumn]{aastex61}

\received{2018 October 29}
\accepted{2018 December 14}
\published{2019 January 3}

\shorttitle{Relic Star Cluster in Sextans}
\shortauthors{Kim et al.}

\begin{document}

\title{A possible relic star cluster in the Sextans dwarf galaxy\footnote{Based on data collected at the Subaru Telescope, which is operated by the National Astronomical Observatory of Japan.}}

\email{sjyoon0691@yonsei.ac.kr}

\author{Hak-Sub Kim}
\affil{Korea Astronomy and Space Science Institute, 776 Daedeokdae-ro, Yuseong-gu, Daejeon 34055, Republic of Korea}

\author{Sang-Il Han}
\affil{Korea Astronomy and Space Science Institute, 776 Daedeokdae-ro, Yuseong-gu, Daejeon 34055, Republic of Korea}
\affil{Department of Astronomy and Space Science, Chungnam National University, 99 Daehak-ro, Daejeon 34134, Republic of Korea}
\affil{Research Institute of Natural Sciences, Chungnam National University, 99 Daehak-ro, Daejeon 34134, Republic of Korea}

\author{Seok-Joo Joo}
\affil{Korea Astronomy and Space Science Institute, 776 Daedeokdae-ro, Yuseong-gu, Daejeon 34055, Republic of Korea}
\affil{Department of Astronomy and Space Science, Chungnam National University, 99 Daehak-ro, Daejeon 34134, Republic of Korea}
\affil{Research Institute of Natural Sciences, Chungnam National University, 99 Daehak-ro, Daejeon 34134, Republic of Korea}

\author{Hyunjin Jeong}
\affil{Korea Astronomy and Space Science Institute, 776 Daedeokdae-ro, Yuseong-gu, Daejeon 34055, Republic of Korea}

\author{Suk-Jin Yoon}
\affil{Department of Astronomy, Yonsei University, Seoul 03722, Republic of Korea}
\affil{Center for Galaxy Evolution Research, Yonsei University, Seoul 03722, Republic of Korea}

\begin{abstract}

We report a possible discovery of a relic star cluster in the Sextans dwarf spheroidal galaxy. 
Using the \textit{hk} index ($\equiv$\,($Ca-b$)\,$-$\,($b-y$)) 
as a photometric metallicity indicator, we discriminate
the metal-poor and metal-rich stars in the galaxy 
and find unexpected number density excess of metal-poor stars located 7$\arcmin$.7 ($\sim$\,190 pc in projected distance) away from the known galactic center.  
The $V-I$ color$-$magnitude diagram (CMD) for stars around the density excess reveals 
that both the main sequence and the giant branch are considerably narrower and redder than the bulk of field stars in Sextans.
Our stellar population models show 
(a) that the narrow CMD is best reproduced by a simple stellar population with an age of $\sim$\,13 Gyr and [Fe/H] of $\sim$\,$-$2.3 dex, and 
(b) that the redder $V-I$ color of the $hk$-weak population
is explained $only$ if it is $\sim$\,2 Gyr older than the field stars.
The results lead us to conclude that the off-centered density peak is likely associated with an old, metal-poor globular cluster.
The larger spatial extent ($>$ 80 pc in radius) and the smaller number of stars ($\sim$\,1000) than typical globular clusters point to a star cluster that is in the process of dissolution. 
The finding serves as the first detection of a surviving star cluster in Sextans, 
supporting previous suggestions of the presence of star cluster remnants in the galaxy.
If confirmed, the survival of a relic star cluster until now implies a $cored$ dark matter halo profile for this dwarf galaxy.
\end{abstract}


\section{Introduction} \label{sec:intro}

In the $\Lambda$ cold dark matter paradigm,
giant galaxies grow by repeated mergers of smaller systems.
Dwarf galaxies, the most common type of galaxies in the universe,
are probably the closest approximation to the building blocks of
larger galaxies.
In particular, dwarf spheroidal (dSph) galaxies around the Milky Way
provide important clues to understand formation and evolution of galaxies,
because their proximity enables us to spatially resolve the galaxies
into individual stars.

The Sextans dSph at a distance of $\sim$\,86 kpc is one of the 
recently discovered satellite galaxies of the Milky Way \citep{Irw90}
with a very low surface brightness ($\mu_{V}$\,=\,28.17 mag arcsec$^{-2}$; \citealt{Mun18})
and unusually large extent on the sky ($r_{h}$\,=\,16$\arcmin$.9; \citealt{Mun18}).
Previous studies have suggested that it
possesses kinematic substructure in the central region.
\citet{Kle04} suggested the presence of a kinematically cold core in the inner five arcmin of the galaxy.
\citet{Wal06}, however, did not confirm the presence of such a core 
but instead detected another kinematically cold substructure north of 
the center at the core radius.
\citet{Bat11} detected similar cold kinematic substructure composed of 
metal-poor stars presumably belonging to a star cluster. 
\citet{Kar12} argued that the chemical properties of the six most metal-poor stars observed by \citet{Aok09}
indicate the presence of a dissolved star cluster within this galaxy.
\citet{Rod16} found an overdense region near the galaxy center 
as well as an extended halolike stellar substructure in the outskirts of Sextans.
\citet{Cic18} detected shell-like stellar overdensity in the spatial distribution
of red stars in $g-r$. They also found a ringlike kinematic substructure 
suggesting a past accretion event.

The survival of stellar clusters in dwarf galaxies 
provides an important constraint on 
the well-known core$-$cusp problem
of the inner dark matter (DM) halo profiles
\citep[for a review, see][]{deB10}.
\citet{Kle03} argued that the two stellar clumps discovered in the Ursa Minor dSph
\citep{Ols85, Irw95, Bel02, Pal03} 
in conjunction with a kinematically cold signature
are not compatible with a cuspy DM halo 
but are consistent with a cored DM halo. 
\citet{Lor13} showed using $N$-body simulations 
that a stellar clump such as a star cluster in Sextans 
can last for a Hubble time preferably in a cored DM halo 
rather than in a cuspy Navarro-Frenk-White \citep{Nav96, Nav97} DM halo.

In this Letter, we investigate the chemostructural properties 
of stellar populations in Sextans 
using a photometric metallicity indicator, the \textit{hk} index.
We discover a spatial overdensity of metal-poor stars
slightly away from the center of the galaxy. 
By comparing the stellar population models with the observations, 
we suggest that the overdensity is associated with 
an old, metal-poor globular cluster 
in the process of dissolution.
We discuss the implications of our results for the shape of DM halo profiles.

\section{Observations and Data reduction} \label{sec:data}

We observed the Sextans dSph galaxy 
with Suprime-Cam on SUBARU 8.2\,m telescope
through $Ca$-, $b$-, and $y$-band filters. 
We also reanalyzed $V$- and $I$-band archival data \citep{Oka17}.  
The standard data reduction procedures were
performed with the SDFRED/SDFRED2 packages \citep{Yag02, Ouc04},
and the photometry was carried out with the DAOPHOT\,II/ALLFRAME packages \citep{Ste87, Ste94}. 
Galactic reddening was corrected with the IRAS dust map \citep{Sch11},
and astrometry was done with respect to the USNO-B\,1.0 catalog \citep{Mon03}. 
Photometric calibration for  $Ca$, $b$, and $y$ bands was performed using the M5 and M15 globular clusters,
and $V$- and $I$-band photometry was calibrated against the standard star catalog provided by \citet{Ste00}.
Details of data reduction and photometry are described in S.-I. Han et al. (2019, in preparation)
that present the detailed chemostructural study on the Draco, Sextans, and Canes Venatici\,\rm{I} dSph galaxies.

\begin{figure*}[ht!]
\epsscale{1.18}
\plotone{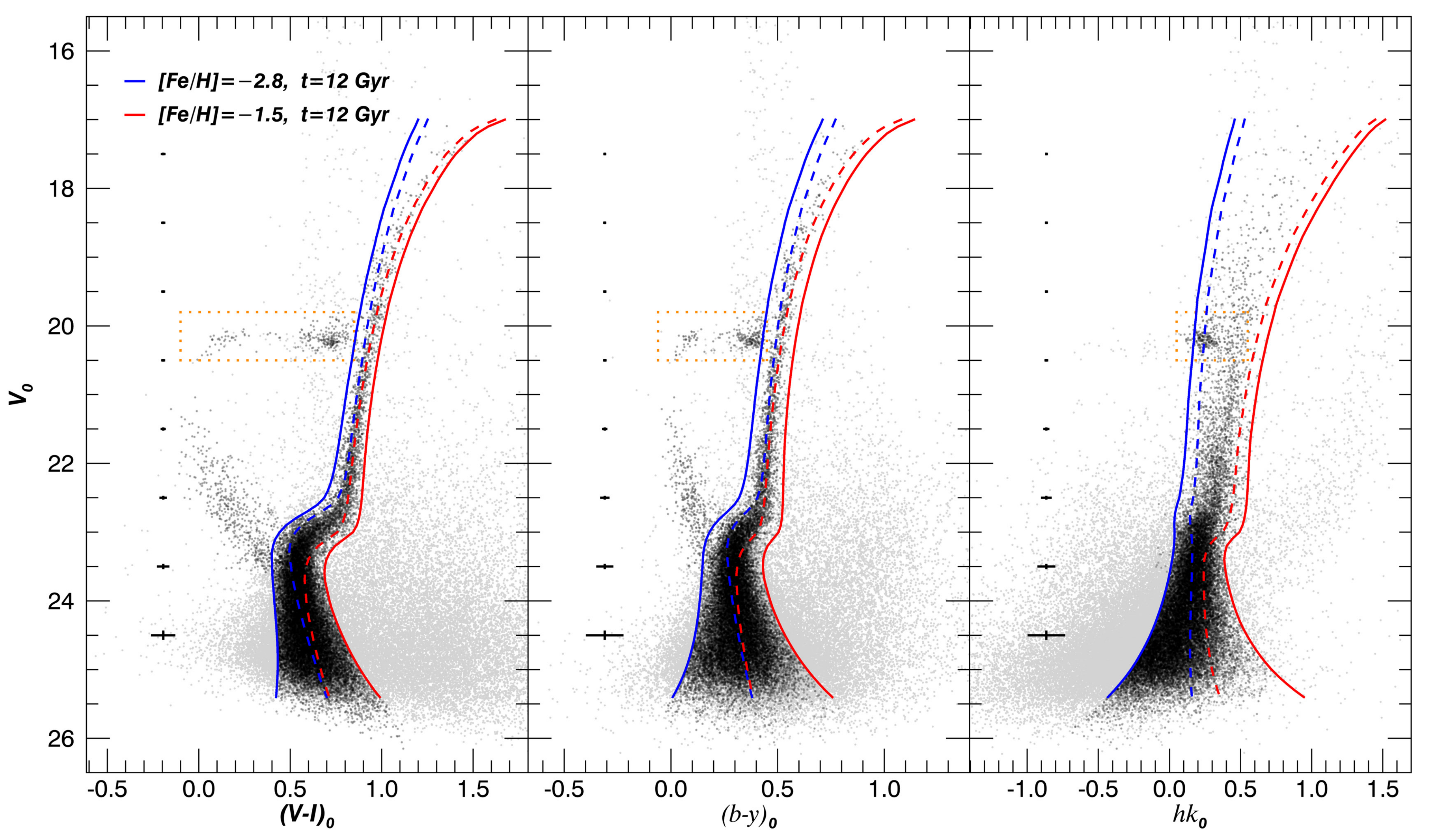}
\caption{Color$-$magnitude diagrams (CMDs) of the Sextans dSph. Blue and red dashed lines represent
Yonsei-Yale isochrones of 12\,Gyr for [Fe/H]\,=\,$-$2.8 and $-$1.5, respectively. 
Blue and red solid lines are modified loci taking into account the observational errors 
for the member selection. The orange dotted line box is the selection area for horizontal-branch stars.
Gray dots are all detected point sources, 
and black dots are selected member stars. 
\label{fig:f1}}
\end{figure*}

\section{Analysis and Results} \label{sec:result}
\subsection{The \rm{hk} \it{index as a photometric metallicity indicator}} \label{subsec:hk index}

The \textit{hk} index, defined as ($Ca-b$)\,$-$\,($b-y$),
is a photometric metallicity indicator using
the $Ca$ filter centered on the ionized calcium $H$ and $K$ lines, 
combined with the str{\"o}mgren $b$- and $y$-photometric system. 
The index was first developed by \citet{Ant91}
by replacing the $v$ filter with the $Ca$ filter from the str{\"o}mgren $m1$ index 
to overcome the low sensitivity of the $m1$ index for metal-poor dwarfs and giants.
The calcium $H$ and $K$ lines remain strong even at very low stellar metallicity
for which other metal lines are quite weak. 
Hence, the \textit{hk} index is an optimal photometric metallicity indicator
for individual stars in dSph galaxies that are normally of low metallicities.

\subsection{Member star selection} \label{subsec:memsel}  

Figure~\ref{fig:f1} shows the color$-$magnitude diagrams (CMDs) of point sources in the area of Sextans.
We select point sources using the ALLFRAME parameters
CHI and SHARPNESS and apply an error cut of 0.2 mag in all bands.
Among the sources (gray dots), we select candidate member stars (black dots)
using theoretical isochrone lines 
for main-sequence (MS) to red-giant-branch (RGB) stars.   
The blue and red dashed lines are the Yonsei$-$Yale (Y$^{2}$) isochrones \citep{Yi08}
of 12\,Gyr for [Fe/H]\,=\,$-$2.8 and $-$1.5, respectively.\footnote{We use 
the synthetic model atmospheres of \citet{Cas03} for the $Ca$, $b$, and $y$ filters (see Section 2.1 in \citealt{Joo13}).}
The metallicity range of the isochrones is determined considering 
the spectroscopic metallicity distributions of RGB stars by \citet{Kir10}.
We modify the isochrones (solid lines) 
taking into account the observational uncertainties 
in order not to miss stars with relatively large errors. 
We select the sources located between the two modified loci in all three CMDs
as the member stars of Sextans.
The horizontal-branch (HB) stars are selected within the regions 
denoted by orange dotted line boxes in all three CMDs. 
Note that we exploit the MS, sub-giant-branch (SGB), RGB, and HB stars in our analysis (see Section~\ref{sec:mod}), 
but we do not use variable stars (e.g., RR Lyrae variable) and blue-straggler (BS) stars.

The use of three colors (i.e., $V-I$, $b-y$, and \textit{hk}) at the same time in the selection of members
significantly reduces contamination by 
the foreground stars and background galaxies.
In particular, using the \textit{hk} index 
efficiently reduces the contamination by foreground stars. 
The RGB stars in the \textit{hk} CMD
are detached from the Galactic disk stars 
that show a strong \textit{hk} index (i.e., metal-rich), 
while they are indistinguishable in optical CMDs. 
The faint member stars ($V$\,$>$\,24.0 mag) are marginally separated from 
the background galaxies, which have relatively redder colors in $V-I$ and $b-y$, 
but have weaker \textit{hk} indices.

\begin{figure*}
\epsscale{1.18}
\plotone{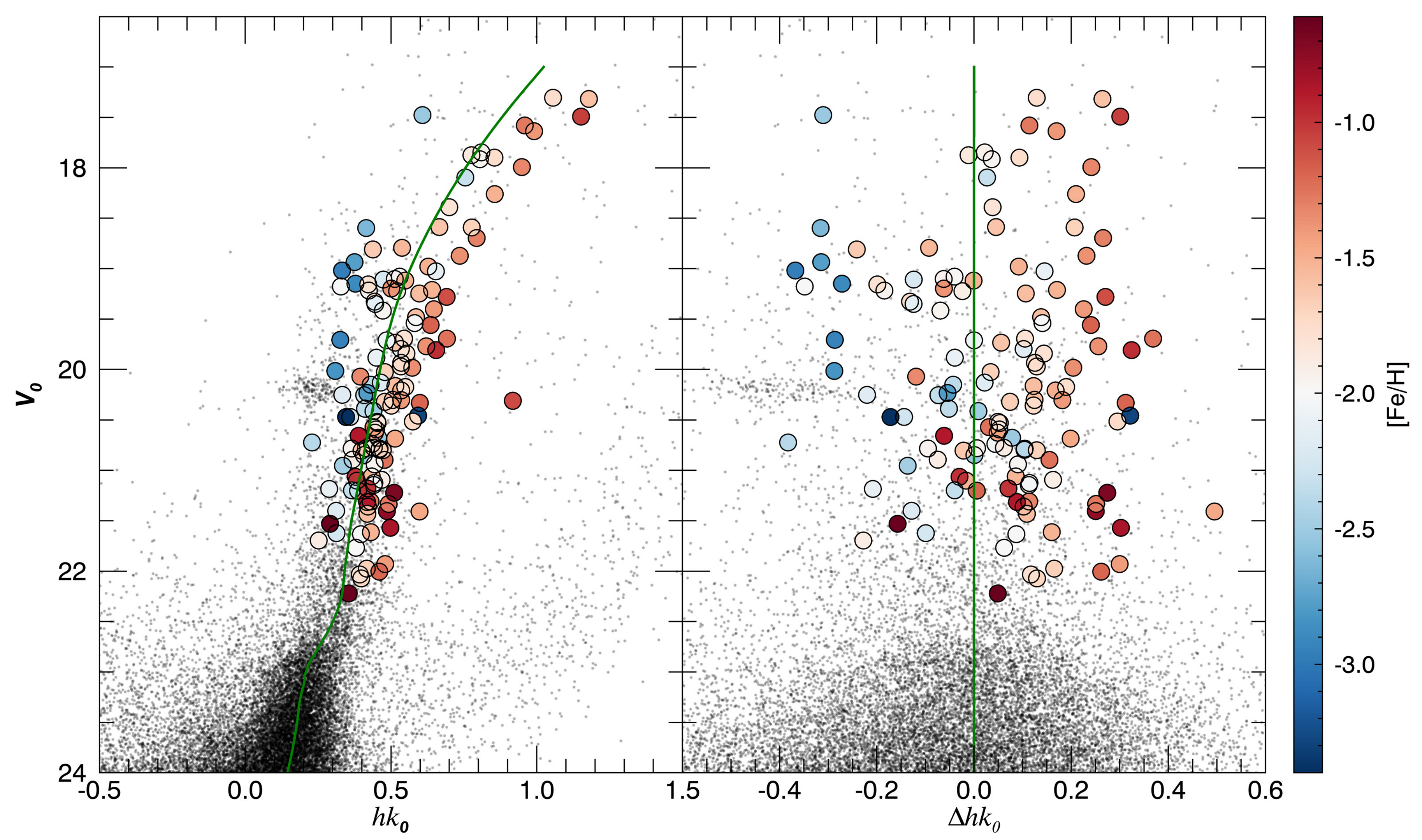}
\caption{The \textit{hk} (left) and verticalized $\Delta$\textit{hk} (right) CMDs 
along with a spectroscopic sample from Kirby et al. (2010). The cross-matched objects 
are color-coded by spectroscopic metallicity shown as a color bar on the right side. 
The green line in the left panel represents the LOESS regression fit to the candidate member stars. 
The $\Delta$\textit{hk} in the right panel is defined as the distance from the fiducial line 
normalized by the \textit{hk} difference between the two isochrones 
used in the member selection. 
\label{fig:f2}}
\end{figure*}

\begin{figure*}
\epsscale{1.18}
\plotone{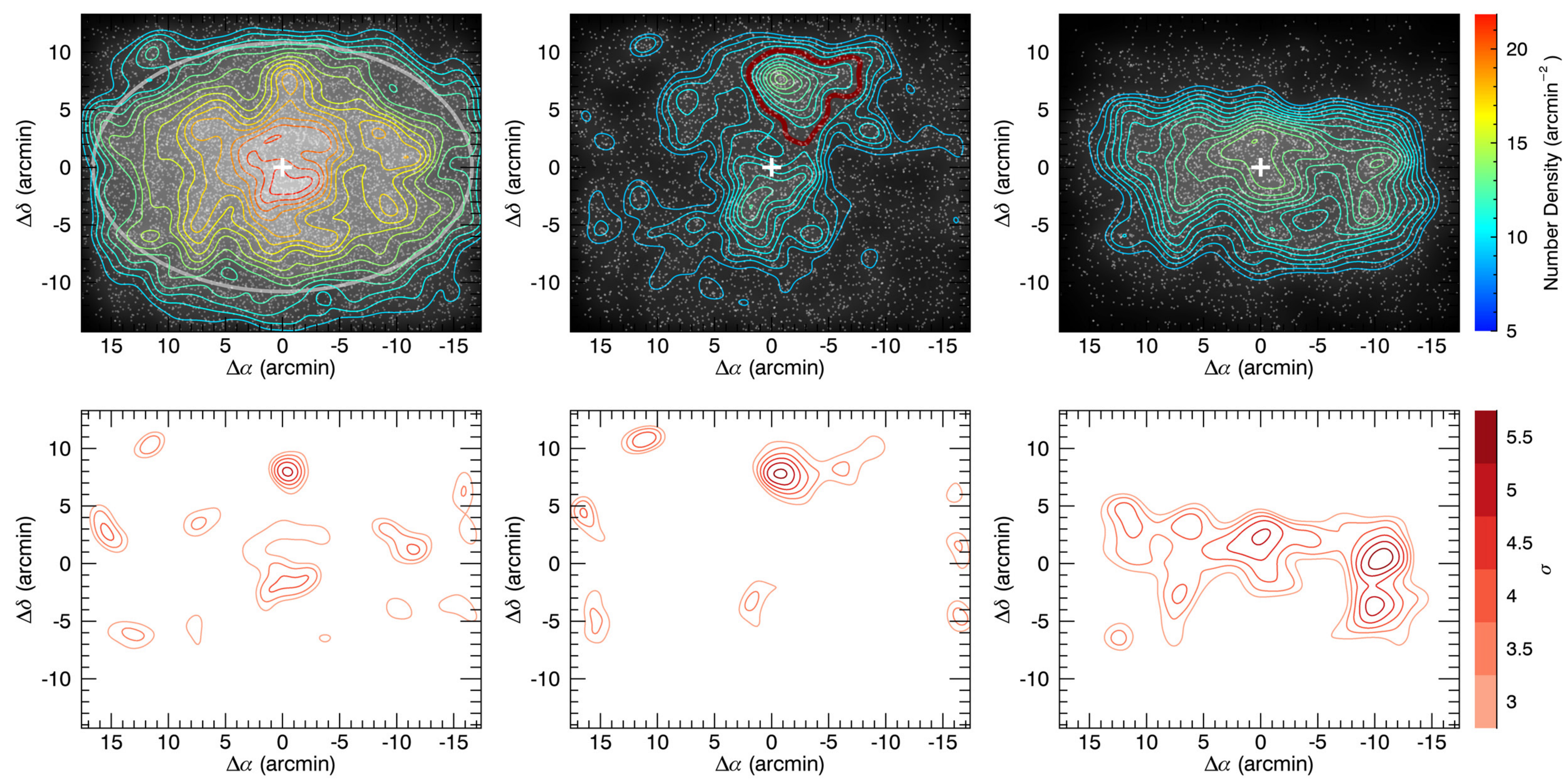}
\caption{Upper row: spatial distributions of all member stars (left), metal-poor stars (center), and metal-rich stars (right) brighter than $V$=\,24.0 
along with the surface number density contours from 
the two-dimensional Kernel density estimation 
using the Sheather \& Jones (1991) bandwidth. 
The images are rotated so that the major axis of Sextans is along the $x$-axis. 
North is to the top left, and east is to the bottom left.
The white cross represents the center of the galaxy \citep{Irw90},
and the gray ellipse denotes the core radius and ellipticity.
In the central panel, an unexpected number excess of metal-poor stars is visible 
at about 7.7 arcmin northwest from the galaxy center.    
The thick red locus is the isodensity contour of 7.5 arcmin$^{-2}$ 
and the selection area for stars 
associated with a possible relic star cluster (see the text).
Lower row: statistical significance contours 
from 3$\sigma$ in steps of 0.5$\sigma$. 
The significance is estimated from bootstrap tests 
with 1000 replicate resamplings for each group.
The off-centered peak of the metal-poor stars visible in the upper central panel 
is statistically significant ($\sim$5.5$\sigma$). 
\label{fig:f3}}
\end{figure*}

\subsection{Dividing stars into two metallicity groups}
Figure~\ref{fig:f2} shows the \textit{hk} (left) and verticalized $\Delta$\textit{hk} (right) CMDs
along with a spectroscopic sample from \citet{Kir10}. The cross-matched objects are 
color-coded by the spectroscopic metallicities shown as the color bar. 
The green solid line in the left panel represents the fiducial line to the MS-to-RGB stars 
obtained by LOESS regression, a nonparametric locally weighted regression \citep{Cle94}.
The $\Delta$\textit{hk} in the right panel is defined as 
\begin{equation}
\textit{$\Delta$hk} \equiv {\textit{hk}_{\rm{\,star}} - \textit{hk}_{\rm{\,fiducial}}\over{\rm{width}}} ,
\end{equation}
where the width is the \textit{hk} difference at the same $V$ magnitude of a star 
between the two modified isochrones used in the member selection.
This figure verifies that the \textit{hk} index is a good metallicity discriminator.

We divide the MS, SGB, and RGB stars into metal-poor and metal-rich groups 
at \textit{$\Delta$hk} = 0. For the MS stars, we use the upper part ($V<$~24.0 mag) only 
because the two groups are mixed in lower MS due to the observational errors.

\subsection{Spatial distributions} \label{subsec: }

Figure~\ref{fig:f3} (upper row) shows the spatial distributions 
of all member stars (left), metal-poor stars (center), and 
metal-rich stars (right) in Sextans.
We construct a surface density map for each group 
shown as a grayscale image and a set of contours 
using the two-dimensional kernel density estimation (KDE) technique \citep{Sil86, Gou14, Sel16}
implemented in the statistical package R\footnote{\url{https://www.r-project.org}}.
The KDE technique is a nonparametric density estimation method,      
which smoothes points using a kernel function 
to infer the underlying probability density distribution of the data.
We use the bivariate normal kernel with the bandwidth
estimated by the method of \citet{She91}.

Comparing the distributions of the two groups 
shows that the metal-rich stars (upper right) are more centrally concentrated,
with an elongation along the major axis of the galaxy.
Overdense regions are visible on the both sides of the densest region,
which seems consistent with the 'shell-like' feature reported by \citet{Cic18}.
On the other hand, the metal-poor stars (upper middle) are relatively dispersed, 
with an elongation rather perpendicular to the major axis direction.

Remarkably, we find an unexpected feature 
in the spatial distribution of metal-poor stars
(a thick red contour in the upper central panel).
Besides a peak at the galaxy center, 
there is another peak at 7.7\,arcmin ($\sim$\,190\,pc) northwest from the center. 
The off-center peak is quite separated from the galactic central density peak. 
The strong concentration of the off-center peak with a roundish shape may point to gravitational boundness. 
The number density at the off-center peak is even higher  
than the galactic central peak among the metal-poor stars. 
We stress that the off-center peak is only prominent among the metal-poor stars,
and when considering the entire populations (upper left), 
the peak is less conspicuous. 
This may be the reason why the structure has not been discovered so far.

Figure~\ref{fig:f3} (lower row) shows the statistical significance contours 
of the number excess for each group.
To quantify the significance of the off-center peak of metal-poor stars, 
we first construct a smoothed surface density map 
using the KDE technique with five-times larger bandwidth 
than that used for constructing the original surface density map, 
and then subtract it from the original map to obtain a residual map. 
We estimate the 68\,\% confidence level (1$\sigma$) of the residual map
from the bootstrap tests with 1000 replicate resamplings\footnote{
Pseudo data sets are generated by sampling the same number of the observed data
with replacement from the original data set.} for each group.
The significance contours show that our off-center peak of metal-poor stars
is detected with a over 5$\sigma$ significance.
The off-center peak is statistically outstanding so that it is the most distinct feature 
in the contour plot of all members (lower left).

\begin{figure*}
\epsscale{1.0}
\plotone{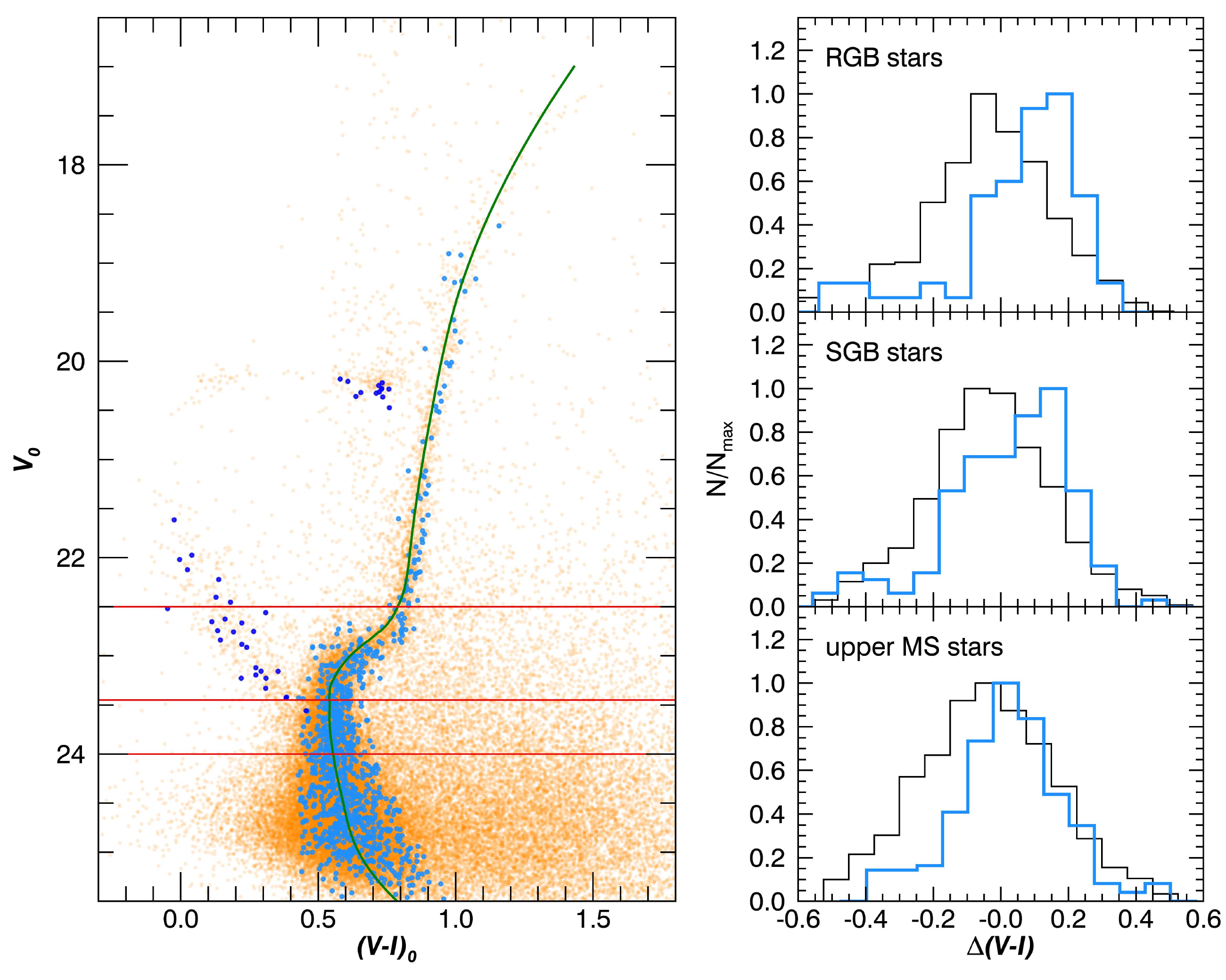}
\caption{
Left: $V-I$ CMD of metal-poor stars around the off-center peak (sky blue dots) 
overlaid on the CMD of all point sources (orange dots). 
The green solid line represents the LOESS regression fit, 
and the red solid lines are the border lines between RGB, SGB, and MS stars. 
Right: $\Delta(V-I)$ distributions of the metal-poor stars around the off-center peak (sky blue) 
and the bulk of field stars in Sextans (black). 
The $\Delta(V-I)$ value indicates the color difference from the LOESS regression fit
where the larger value means the redder color. 
The height of the histogram of each group is normalized to unity for easier comparison.
\label{fig:f4}}
\end{figure*}

\begin{figure*}
\epsscale{0.9}
\plotone{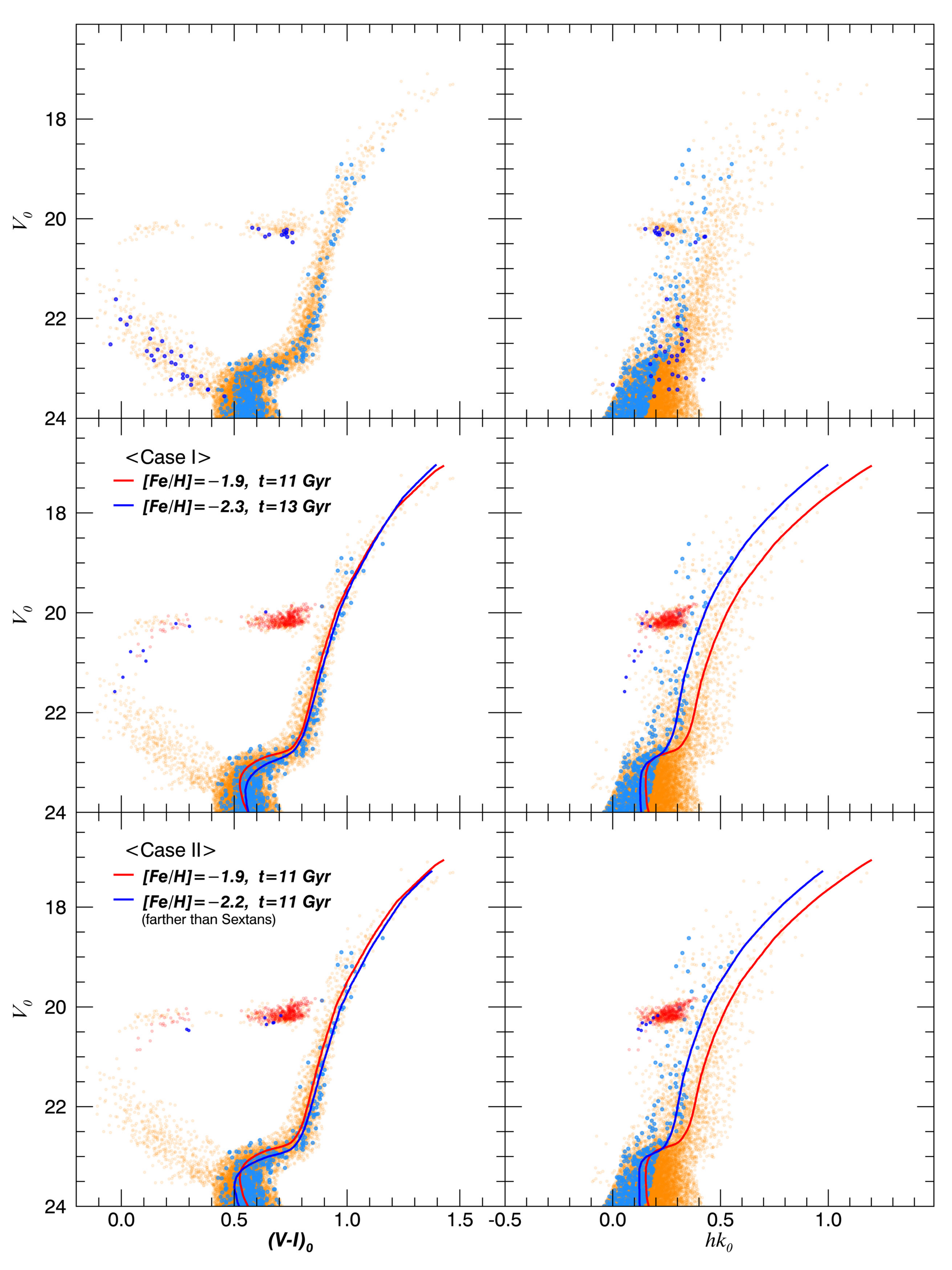}
\caption{Top: $V-I$ (left) and \textit{hk} (right) CMDs 
of the metal-poor stars around the off-center peak (sky blue dots) 
overlaid with the bulk of field stars in Sextans (orange dots). 
All of the HB and BS stars around the peak,
metallicities for which cannot be determined by the $hk$ index, are shown as well. 
Middle: the ``age" model. Theoretical model $V-I$ (left) and
\textit{hk} (right) CMDs are overlaid with the observations (orange dots). 
The red and blue colors are for the isochrones and synthetic HB stars 
of 11\,Gyr for [Fe/H]\,=\,$-$1.9 and 13\,Gyr for [Fe/H]\,=\,$-$2.3, respectively. 
Bottom: the ``distance" model. 
The red and blue colors are for the 11 Gyr isochrones and synthetic HB stars for [Fe/H]\,=\,$-$1.9 and [Fe/H]\,=\,$-$2.2, respectively. 
In this model, the metal-poor stars around the off-centered peak 
are set to be fainter by 0.25 mag in $V$ than the observed field stars of Sextans, 
by assuming the former is $\sim$\,10.5\,kpc farther than the latter.
\label{fig:f5}}
\end{figure*}

\subsection{Properties of the stars around the off-center peak: A possible relic star cluster}
\label{sec:mod}

Figure~\ref{fig:f4} shows 
the $V-I$ CMD (left) and the color distributions (right) 
of the metal-poor stars around the off-centered density peak, 
compared with those of the bulk of field stars in Sextans. 
In the CMD, a total of 1035 metal-poor stars 
selected in the off-center peak region 
are overlaid with all point sources. 
We also plot the all of the HB and BS stars detected in this region. 
For HB stars, the $hk$-based metallicities cannot be determined in this study.

The narrow sequence defined by the stars of the off-center peak region 
in the left panel suggests a possible association with a single star cluster.
In the right panel, we compare the color distributions of RGB, 
SGB, and upper MS 
between the selected stars around the off-center peak
and all member stars in Sextans. 
The $\Delta(V-I)$ value is the difference between a star's color and 
the LOESS regression fit (green solid line).
The histogram shows that the stars exhibit redder $V-I$ colors 
than those of the bulk stars in Sextans.
This phenomenon is very puzzling, 
given that the stars around the off-center peak 
are $hk$-weak, metal-poor stars and thus expected to have bluer $V-I$ colors.

To address this puzzling phenomenon, 
we attempt to reproduce the observations with theoretical population models
following the technique developed by \citet{Yoo08} and \citet{Joo13}. 
We use the Y$^{2}$ isochrone and HB evolutionary tracks (\citealt{Yi08, Han09}).
In the top panels of Figure~\ref{fig:f5}, we present 
the observed $V-I$ and \textit{hk} CMDs, 
where the sky blue dots are the stars around the off-center peak 
and the orange dots are all the member stars in Sextans. 
Blue dots represent all the HB and BS stars detected around the off-center peak.
We first assume that, despite the internal metallicity and age spread of the galaxy, 
Sextans's CMD can be roughly represented by a simple stellar population. 
We adopt a metallicity of [Fe/H]\,=\,$-$1.9 and an age of 11\,Gyr\footnote{Recent 
determination of Sextans star formation history (Bettinelli et al. 2018) shows that 
over 80\,\% of stars are older than 12\,Gyr. But we note that a careful approach is 
required to compare the absolute age values derived from CMDs because they
depend fairly on the population model used.} in our model. 
The metallicity is from the spectroscopic metallicity distribution of RGB stars \citep{Bat11, Kir11}
and the age is adjusted until we obtain the best match
with the observations (red solid lines in the middle and bottom panels).
We then seek a population representing the off-center peak stars, 
which have weaker \textit{hk} strength but redder $V-I$ color 
than the bulk of stars in Sextans. 
We test two possible scenarios for the \textit{reversal} of the $V-I$ color.

In the first case, 
we assume an old, metal-poor star cluster population for 
the off-center peak (blue solid lines in the middle panels). 
Compared to the main body of Sextans, 
the lower metallicity can explain the weak $hk$ strength 
while the younger age by 2\,Gyr can naturally bring about the redder color 
of metal-poor stars in $V-I$: the median $\Delta(V-I)$ difference between 
the two groups in the observations is $\sim$\,0.080 and the reproduced value is $\sim$\,0.067.
The metallicity and age differences are reasonable, 
considering a wide metallicity spread \citep{Bat11, Kir11}
and a large age range ($>$\,3\,Gyr; \citealt{Lee09, Oka17}) in this galaxy.  
This ``age" model, however, predicts several blue HB stars (blue dots)
for the star cluster population. 
Such blue HB stars are not observed within the off-center peak region 
but rather some red HB stars are present.
Given that the small number of RGB stars ($<$\,65) of the star cluster population,
the expected number of HB stars are as small as $<$\,3. 
Recall that we cannot determine HB stars' metallicities based on the $hk$ index.
We thus suspect that many of 13 red HB stars observed around the off-center peak 
actually belong to the metal-rich population of Sextans.
We hence propose that the off-centered metal-poor peak 
is associated with a relic star cluster.

In the second case, 
we alternatively assume a metal-poor star cluster
that is farther than the observed stars in the vicinity of Sextans's center (blue solid lines in the lower panels).
An observational hint for the distance difference 
comes from the $V$ magnitude distributions of HB stars, 
where the HB stars around the off-center peak
appear to be slightly fainter than the rest of the stars.
The ``distance" model, with a younger age of 11\,Gyr, 
is able to reproduce observed red HB stars. 
The larger distance ($\sim$\,10\,kpc farther than Sextans) 
makes the RGB and SGB sequence 
fainter than those of Sextans's stars, resulting in the redder $V-I$ colors. 
However, the large galactocentric distance of the star cluster 
from Sextans
casts doubt on the cluster being bound to Sextans.
Rather it could be an extended globular cluster in the Milky Way's outer halo 
($d$\,$\sim$\,100 kpc) like Palomar 14 \citep{Sol11},  
which is placed in the direction of Sextans by chance.
Although this is an interesting possibility, 
the model, nevertheless, inevitably predicts a bluer color of the star cluster 
at the MS compared to that of Sextans's stars, 
in contrast to the observation. 
It also seems difficult to explain the distorted feature
observed near the off-center peak.
Hence, it seems that the ``distance" scenario is implausible, 
although it awaits further investigation.

\section{Summary and Discussion}

We examined the chemostructural properties of the Sextans dSph 
by dividing the member stars into metal-poor and metal-rich groups
using the photometric metallicity indicator, \textit{hk} index.
From the spatial distribution of metal-poor stars, 
we found an unexpected number density peak
at about 7.7\,arcmin ($\sim$\,190\,pc) northwest from the center of the galaxy.
We proposed based on the CMD analysis that the off-centered metal-poor peak 
is associated with a metal-poor star cluster, 
which is older than the rest of observed stars in Sextans.
The large size of the roundish overdense region ($\sim$\,80\,pc in radius)
and the smaller number of stars ($\sim$\,1000) compared to typical globular clusters
may imply that the cluster is under dissolution but not completely dissolved yet. 
This is consistent with previous studies suggesting the presence of star cluster remnants in this galaxy
on the basis of kinematically cold substructures \citep{Kle04, Wal06, Bat11} 
or chemical coherence of several stars \citep{Kar12}, 
although our star cluster does not coincide in position with them.

The survival of a star cluster within Sextans
provides an important clue for the core$-$cusp problem.
Several numerical simulations \citep[e.g.,][]{Kle03, Lor13} 
suggest that star clusters can persist for a Hubble time preferentially 
in the cored DM halo than in the cuspy DM halo. 
\citet{Ass13a, Ass13b} predicted, in their dwarf galaxy formation model, 
that, in a cored DM halo, some of star clusters can survive through the Hubble time 
if their apocentric distance is close enough ($<$\,300\,pc; see also \citealt{Ala18}) to reside 
in the core region where little potential gradient is present.
The star cluster we identified has a projected distance of $\sim$\,190\,pc 
from the galaxy center, and this could explain the reason why the old cluster 
was not completely dissolved yet.
Our results thus provide a piece of evidence in favor of the cored DM halo profile of this dwarf galaxy.

Finally, it is worth to mentioning that while we interpret the off-centered overdensity 
of the metal-poor stars as a dissolving globular cluster, other interpretations are possible; it may be, for instance, an ultra-faint dwarf galaxy. 
There are three large spectroscopic datasets 
covering the RGB stars in Sextans by \citet{Wal09}, \citet{Kir10}, and \citet{Bat11}. 
We have tried to cross-match the stars around the off-centered peak with these datasets. 
Unfortunately, the number of cross-matched stars is too small ($\leq$\,9) for a statistically 
significant study. Further spectroscopic study is required to clarify 
the nature of the off-centered metal-poor overdensity.

\acknowledgments
S.-J.Y. acknowledges support from the Center for Galaxy
Evolution Research (No. 2010-0027910) through the NRF of
Korea and from the Yonsei University Observatory$-$KASI Joint
Research Program (2018). H.J. acknowledges support from the Basic Science Research Program through the NRF of Korea, funded by the Ministry of Education (NRF-2013R1A6A3A04064993).



\end{document}